\documentclass[11pt]{article}

\usepackage[preprint]{acl}
\usepackage{times}
\usepackage{latexsym}
\usepackage[T1]{fontenc}
\usepackage[utf8]{inputenc}
\usepackage{microtype}
\usepackage{inconsolata}
\usepackage{booktabs}
\usepackage{amsmath}
\usepackage{amsfonts}
\usepackage{array}
\usepackage{graphicx}
\usepackage[percent]{overpic}
\usepackage[svgnames]{xcolor}
\usepackage{enumitem}
\usepackage[most]{tcolorbox}
\usepackage{placeins}

\newcommand{\guidance}{latent PRM guidance}
\definecolor{promptborder}{RGB}{58,74,91}
\definecolor{promptbg}{RGB}{248,250,252}
\definecolor{prompttitlebg}{RGB}{231,237,244}
\newtcolorbox{promptbox}[1]{
    enhanced,
    breakable,
    colback=promptbg,
    colframe=promptborder,
    colbacktitle=prompttitlebg,
    coltitle=black,
    fonttitle=\bfseries,
    title={#1},
    boxrule=0.6pt,
    arc=1.5mm,
    left=1.2mm,
    right=1.2mm,
    top=0.9mm,
    bottom=0.9mm,
    attach boxed title to top left={xshift=1.2mm,yshift*=-\tcboxedtitleheight/2},
    boxed title style={boxrule=0.6pt,arc=1mm,colframe=promptborder},
    before skip=0.75em,
    after skip=0.75em
}
\newcommand{\promptplaceholder}[1]{\texttt{#1}}

\title{Latent Reasoning Guidance for Parallel Code Translation}
\author{
\textbf{Tomer Bitan}$^{1}$ \quad
\textbf{Erel Kaplan}$^{1}$ \quad
\textbf{Roee Bar-Yadin}$^{1}$ \quad
\textbf{Lian Ghrayeb}$^{1}$ \\
\textbf{Le Chen}$^{2}$ \quad
\textbf{Samyak Jhaveri}$^{3}$ \quad
\textbf{Niranjan Hasabnis}$^{4}$ \quad
\textbf{Gal Oren}$^{1,5}$ \\
$^{1}$Technion \quad
$^{2}$Argonne National Laboratory \quad
$^{3}$University of California, Irvine \\
$^{4}$Code Metal \quad
$^{5}$Stanford University
}
\date{}

\begin{document}
\maketitle

\begin{abstract}
Tackling complex coding tasks often requires autonomous agents and iterative repair pipelines. These increasingly rely on large amounts of test-time computation, often spending many decoding and repair steps before discovering whether a program compiles, runs, or validates.
Executable parallel-code translation is an effective setting for earlier guidance because success is behavioral rather than textual. However, most guidance methods act only after complete programs or textual traces are decoded.
This motivates the question: can latent reasoning provide an earlier intervention point, before the model commits to code?

We study a test-time latent guidance method for this setting that trains a smaller Process Reward Model (PRM) over continuous latent prefixes and uses it to select among alternate hidden-state trajectories before final code decoding, separately from but compatible with post-decoding optimization.
On a 76-task ParaTrans benchmark evaluation, latent PRM guidance improves mean validation rate from 32.89\% with unguided latent reasoning to 42.1\%, outperforming fine-tuned and vanilla baselines in the same setting. These gains persist under the same three-attempt repair loop.
These results provide bounded evidence that useful alternative latent continuations exist and that PRM-scored latent branch selection can improve executable outcomes in this setting without retraining the main generative model.
\footnote{GitHub:\url{https://github.com/Scientific-Computing-Lab/Parallax.git}}
\end{abstract}

\section{Introduction}

Scientific and high-performance applications often need to move between programming models such as CUDA, OpenMP, OpenCL, and serial CPU implementations as hardware, compiler support, and deployment constraints change \citep{kadosh2023quantifying}.
Parallel API translation is therefore an important but difficult code-generation problem because translations must preserve the source computation while satisfying target-specific requirements for parallel structure, memory movement, synchronization, and API usage \citep{10158214, bitan2025unipar}.

Large language models and autonomous code agents can propose flexible transformations beyond rigid source-to-source rules \citep{kadosh2023scopeneedtransformingllms, chen2024landscapechallengeshpcresearch, kaplan2026paracodexprofilingguidedautonomouscoding}, but in executable parallel-code translation, even plausible looking outputs may fail to compile, crash, or violate reference behavior, pushing many systems toward generate$\rightarrow$test$\rightarrow$repair loops.

Small differences in intermediate reasoning can lead to different compile$\rightarrow$run$\rightarrow$validate outcomes, and poor high-level parallelization choices can be hard to repair after decoding. This motivates an earlier intervention point: can nearby reasoning trajectories lead to better executable outcomes before any token-level repair is attempted?

Existing guidance methods typically score decoded programs, textual reasoning traces, or completed candidates \citep{cobbe2021trainingverifierssolvemath, lightman2023let, li2025codeprm, dai2024process}.

Latent reasoning offers an earlier intervention point by shifting some test-time computation into continuous hidden-state trajectories before standard decoding resumes \citep{hao2025traininglargelanguagemodels, xu2025softcot, yue2025hybrid, zou2025latentcollaborationmultiagentsystems}.
An intervention during the continuous phase can have a large effect redirecting subsequent latent steps and decoded tokens before the model commits to an implementation strategy.
This is especially relevant to parallel-code translation, where early choices about work decomposition, memory movement, and synchronization constrain much of the final program.
However, recent work has begun to investigate this intervention point through latent-state optimization \citep{you2026parallel, li2026latentrewardsteering}  We examine this question in executable parallel-code translation, a domain that requires specialized parallel-programming knowledge and in which pre-decoding guidance must remain compatible with post-decoding repair.

We study a test-time latent guidance method for executable parallel API translation that keeps the primary model frozen, samples perturbed hidden-state branches during latent reasoning, scores partial latent prefixes with a smaller PRM, and resumes generation from the highest-scoring branch, placing task-specific adaptation in the compact guidance model rather than retraining the generator.
Related work and positioning are discussed in Appendix~\ref{sec:appendix-related}.

Empirically, \guidance{} improves ParaTrans validation by 9.21 points over unguided latent reasoning, and the gain persists when inserted into the same three-attempt repair loop.
A held-out branch-selection test further suggests that the PRM can learn a weak but useful local preference signal to avoid undesirable latent continuations.

We make three contributions.
First, we study pre-decoding latent branch selection for executable parallel-code translation and test its compatibility with post-decoding repair.
Second, we instantiate this intervention with a smaller PRM and evaluate it on ParaTrans, where correctness requires compilation, execution, and validation rather than textual similarity.
Third, we provide bounded mechanism-level evidence from a held-out branch-selection test that a smaller PRM can provide a conservative local preference signal over continuous latent branches before final code decoding.

\begin{figure*}[t]
    \centering
    \includegraphics[width=1.0\textwidth,trim=0.25cm 0.3cm 0.25cm 0.0cm,clip]{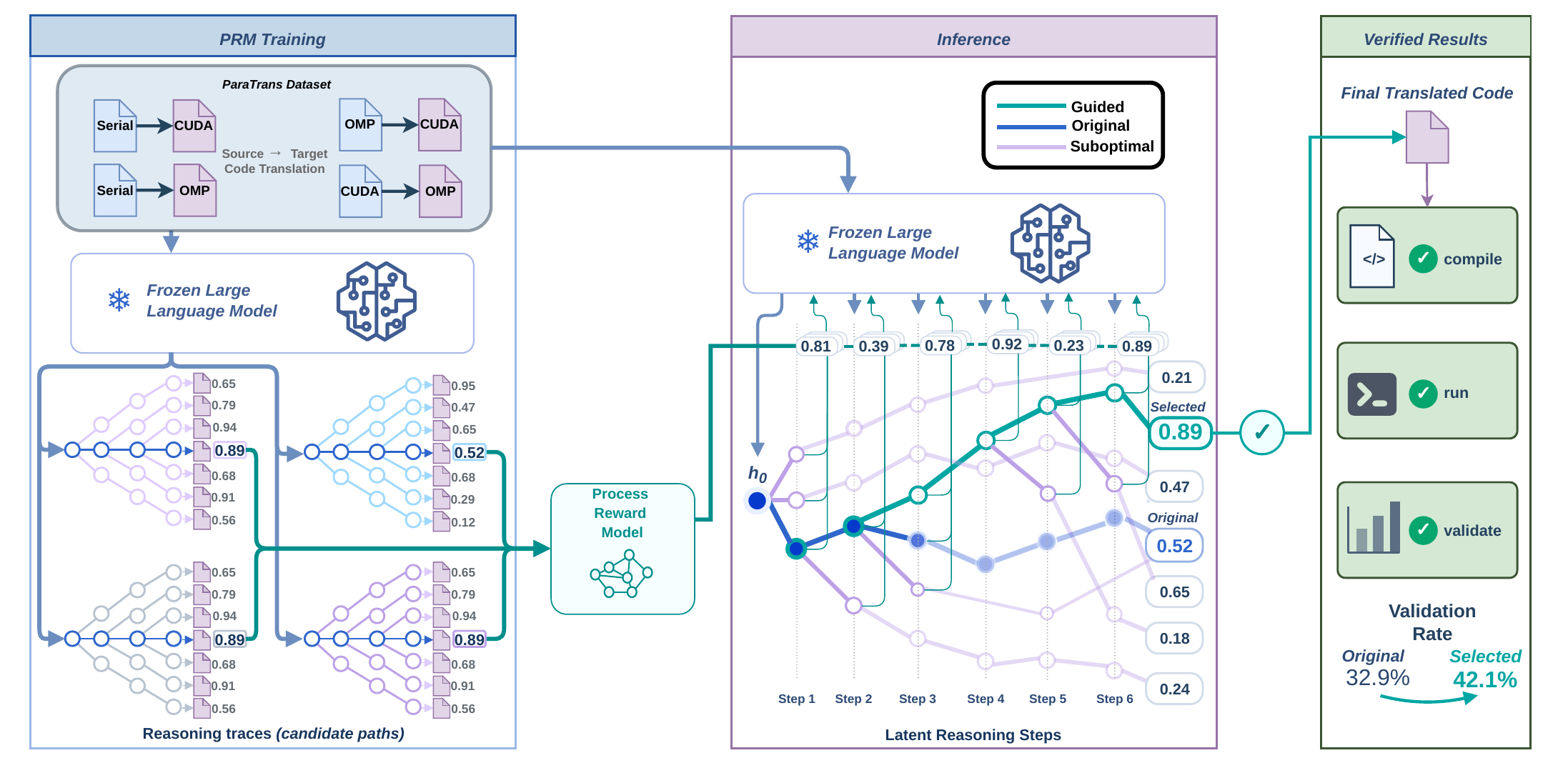}
    \vspace{-0.6em}
\caption{
\textbf{Latent PRM-guided translation pipeline.}
Executable feedback is converted into a pre-decoding guidance signal for parallel-code translation.
(1) PRM training uses ParaTrans paired tasks across Serial, CUDA, and OpenMP: a frozen latent-reasoning generator samples candidate hidden-state paths, decodes them, and assigns supervision from executable outcomes.
(2) The process reward model learns to score partial latent branches before code is emitted.
(3) During inference, a new source task is processed by the frozen generator; candidate latent branches are proposed at each step, scored by the PRM, and reduced to the highest-scoring trajectory before final decoding.
(4) Verified results are obtained by compiling, running, and validating the translated code against the reference behavior.
This design tests whether latent trajectories can be steered before decoding while leaving the main generator frozen.
}

    \label{fig:parallel-prm}
\end{figure*}

\section{Method}
\label{sec:method}

We instantiate a test-time optimization method for guiding latent reasoning without fine-tuning the main generator in parallel code translation.
Specifically, we keep a large latent-reasoning primary model frozen and train a smaller PRM to score partial latent trajectories before text decoding, as shown in Figure~\ref{fig:parallel-prm}.

\subsection{Latent-reasoning Primary Model}

Let $M$ be a frozen autoregressive model equipped with continuous latent reasoning.
In our main experiments, $M$ is LLaMA-3.3-70B.
Following the COCONUT-style latent-reasoning paradigm \citep{hao2025traininglargelanguagemodels}, the model performs a fixed number of latent steps before returning to standard token decoding.
Let $h_t \in \mathbb{R}^{d_M}$ denote the primary model's hidden state at latent step $t$, before the language-modeling head.
During the latent phase, $h_t$ is not decoded into a discrete token; instead, it is mapped back into the $M$ input space and used as the next continuous input.

Instead of training the main model in the manner proposed in \citet{hao2025traininglargelanguagemodels}, we use the training-free alignment transform of \citet{zou2025latentcollaborationmultiagentsystems}.
Given the input embedding matrix $W_{\mathrm{in}}$ and output unembedding matrix $W_{\mathrm{out}}$, we compute
\[
W_a = (W_{\mathrm{out}}^\top W_{\mathrm{out}})^{-1} W_{\mathrm{out}}^\top W_{\mathrm{in}},
\]
the closed-form solution to
\[
\min_{W_a} \| W_{\mathrm{out}} W_a - W_{\mathrm{in}} \|_F^2 .
\]
At latent step $t$, the aligned continuous input for the next step is $z_{t+1}=h_t W_a$.
We use $K_{\mathrm{train}}=6$ latent steps when collecting PRM training trajectories and $K_{\mathrm{test}}=12$ latent steps during inference.

\subsection{Latent process reward model}

We train a smaller reward model to estimate the potential of a partial latent trajectory of the reasoning process via the mean terminal quality score.
The PRM uses a Qwen-Coder-7B backbone with a scalar value head.
Because the PRM receives hidden states from the primary LLaMA-3.3-70B model rather than its own token embeddings, we prepend a learned adapter that maps this hidden dimension to the PRM embedding dimension.
In our implementation, this maps 8192-dimensional generator states to the 3584-dimensional Qwen-Coder input space using a linear layer, GELU activation, and layer normalization.

For every latent reasoning step $t$, the PRM $V_\theta$ receives the prefix trajectory
\[
\tau_{\leq t} = (h_1,\ldots,h_t)
\]
and predicts a scalar value
\[
V_\theta(\tau_{\leq t}).
\]
The target is the empirical terminal reward obtained from rollout sampling.

For rollout supervision at each latent step $t$, we sample $B_{\mathrm{train}}=3$ candidate branch states: two perturbed resamples and the unperturbed hidden state.

\begin{equation}
\label{eq:perturbation}
\tilde{h}_{t}^{(b)} = h_t + \epsilon^{(b)},
\qquad
\epsilon^{(b)} \sim \mathcal{N}(0, 1.4^2 I).
\end{equation}

For each branch $b$, we roll out to terminal code $y_t^{(b)}$ and score the completed program with an automatic reward function $S(\cdot)$.

The prefix target is the average terminal score of its continuations:
\[
R_t = \frac{1}{B_{\mathrm{train}}}
\sum_{b=1}^{B_{\mathrm{train}}}
S(y_t^{(b)}),
\]

The PRM is trained with mean squared error:
\[
\mathcal{L}
=
\frac{1}{N}
\sum_{i=1}^{N}
\left(
V_\theta(\tau_{\leq t}^{(i)}) - R_t^{(i)}
\right)^2 .
\]

\subsection{Training data and reward construction}

We construct the PRM training set from primary-model reasoning trajectories on ParaTrans samples used by \citet{bitan2025unipar} and supplement with additional samples created with the same methodology, balanced across CUDA$\rightarrow$OpenMP, OpenMP$\rightarrow$CUDA, Serial$\rightarrow$OpenMP, and Serial$\rightarrow$CUDA.
We use 60 development samples for hyperparameter selection and train the final PRM on the combined training and development set before evaluating on the held-out test tasks, as shown on the left side of Figure~\ref{fig:parallel-prm}.

The terminal reward $S(y)$ combines executable and semantic signals, providing denser supervision than executable feedback alone. It is distinct from the reported benchmark metric.
Compilation and execution receive fixed weights of $0.30$ and $0.25$, respectively.
The remaining $0.45$ weight is distributed across the validators available for the sample, proportional to their base weights.
These validators include pass/fail validation, checksum or output matching when available, benchmark metrics, code-embedding similarity \citep{günther2024jinaembeddings28192token}, an LLM-as-a-judge score, and a variable-comparison heuristic inspired by \citet{kaplan2026paracodexprofilingguidedautonomouscoding}.
In contrast, final benchmark results use executable validation only: a translation must compile, run, and pass integrated validation against the reference behavior.
Appendix~\ref{sec:appendix-hparams} gives the exact reward construction and validator details, and Section~\ref{sec:abilation} reports a grouped reward ablation.

\subsection{PRM-guided latent inference}

At inference time, we perform greedy branch selection during the latent reasoning phase, as shown in the inference panel of Figure~\ref{fig:parallel-prm}.
For every latent step $t$, we score $B_{\mathrm{test}}=8$ candidate branch states: the unperturbed state $h_t$ and seven perturbed variants from Eq.~\ref{eq:perturbation}.

Each candidate forms a possible branch
\[
\tau_{\leq t}^{(b)} = (h_1,\ldots,h_{t-1},\tilde{h}_t^{(b)}).
\]
The PRM scores each candidate branch and selects
\[
b_t^\star
=
\arg\max_{b}
V_\theta(\tau_{\leq t}^{(b)}).
\]
Only the selected branch is retained.
The selected hidden state $\tilde{h}_t^{(b_t^\star)}$ is aligned with $W_a$ and fed back into the primary model for the next latent step.
After $K=12$ latent steps, the primary model returns to standard autoregressive decoding and produces the final target code. While initial training is done with $K=6$ latent steps for efficiency, we find that using more latent steps at inference can further improve performance, as shown in Appendix~\ref{sec:appendix-hparams}.

\section{Experimental Setup}
\label{sec:setup}

The main evaluation uses the 76-task ParaTrans test set.
Validation rate is the three-run mean fraction of translations that compile, execute without timing out, and pass all task-specific behavioral checks against the benchmark reference implementation, including output, checksum, or benchmark-specific checks where applicable.
This is stronger than compilation or textual similarity, although it remains benchmark-level behavioral testing rather than a formal proof of semantic equivalence.
We compare vanilla LLaMA-3.3-70B, the fine-tuned LLaMA-3.3-70B baseline from \citet{bitan2025unipar}, unguided latent reasoning with the same frozen base model, random latent branch selection with the same perturbation budget, and \guidance{}. We do not evaluate latent reasoning on the fine-tuned generator or a combined fine-tuned-generator-plus-latent-PRM setting.
Appendix~\ref{sec:appendix-hparams} gives the full evaluation protocol, including direction counts, overlap checks, the repair setting, and the held-out branch-selection benchmark; direction-wise results appear in Appendix~\ref{sec:appendix-direction}.

\section{Results}
\label{sec:results}

\subsection{Main executable translation results}

\begin{table}[t]
    \centering
    \small
    \caption{Three-run mean executable validation rate (\%) on the ParaTrans test set, with and without the same three-attempt repair loop.}    \label{tab:main-results}
    \begin{tabular*}{\columnwidth}{@{\extracolsep{\fill}}lcc}
        \toprule
        Method & No repair & +3-attempt repair \\
        \midrule
        Vanilla & $14.55{\scriptstyle \pm 2.55}$ & $30.67{\scriptstyle \pm 4.34}$ \\
        Fine-tuned & $33.48{\scriptstyle \pm 1.11}$ & $36.00{\scriptstyle \pm 3.11}$ \\
        Latent reasoning & $32.89{\scriptstyle \pm 3.44}$ & $36.40{\scriptstyle \pm 5.47}$ \\
        Random branch selection & $27.28{\scriptstyle \pm 6.34}$ & -- \\
        Latent PRM guidance & $\mathbf{42.10}{\scriptstyle \pm 2.28}$ & $\mathbf{45.18}{\scriptstyle \pm 3.31}$ \\
        \bottomrule
    \end{tabular*}
\end{table}

Table~\ref{tab:main-results} reports executable validation rates on ParaTrans.
The key comparison is between unguided latent reasoning and \guidance{}, since both use the same frozen base model and differ only in whether latent branches are scored and selected.
Latent PRM guidance improves validation by 9.21 percentage points without repair (paired task-bootstrap 95\% CI [+3.07, +15.79], two-sided paired sign-flip $p=0.0081$) and by 8.78 percentage points under the same three-attempt repair loop (95\% CI [+2.63, +15.35], $p=0.0064$).
A random branch-selection baseline with the same perturbation budget reaches only 27.28\% validation, suggesting that many latent perturbations are harmful rather than automatically useful.
Without repair, latent PRM guidance already exceeds all non-PRM baselines, including repaired latent reasoning. With repair, it improves further to 45.18\%, reinforcing the notion that pre-decoding guidance remains useful alongside later token-level repair.
Section ~\ref{sec:abilation} reports the reward and latent-search-budget ablations.
Appendix~\ref{sec:appendix-supporting-results} reports the test-time overhead comparison, task-level win/loss/tie counts, direction-wise results, and a post-decoding best-of-8 control.

\subsection{Latent-trajectory branch analysis}

We test whether perturbed latent continuations can improve on the unperturbed trajectory.
On the 952-sample PRM training set, 95\% of trajectory samples contain at least one candidate branch with higher terminal reward than the original. 59\% of candidate continuations improve over their corresponding unperturbed branch. Among samples with at least one trajectory that passes both validation and output comparison, 25\% succeed only through branching: the original trajectory fails, while at least one perturbed continuation succeeds.
These results show that better latent continuations often exist locally, though distinguishing them from harmful ones remains difficult.

\subsection{Branch-selection analysis}

The branch-selection test isolates whether the PRM can choose a useful latent branch from a fixed local candidate set.
To avoid leakage, this PRM excludes the 120 training trajectories used to construct the 500 branch-selection instances.

\begin{table}[t]
    \centering
    \small
    \caption{Held-out branch-selection results on 500 instances. ``Best'' is strict best-branch accuracy; $>$ Orig. is strict improvement over the original continuation, and $\geq$ Orig. is at least as good as the original.}
    \label{tab:branch-selection}
    \begin{tabular*}{\columnwidth}{@{\extracolsep{\fill}}lccc}
        \toprule
        & \multicolumn{3}{c}{Selection rate (\%)} \\
        \cmidrule(lr){2-4}
        Selector & Best & $>$ Orig. & $\geq$ Orig. \\
        \midrule
        Original & 33.66 & - & -\\
        Random & $32.14{\scriptstyle \pm 0.46}$ & $43.0{\scriptstyle \pm 0.4}$ & $66.9{\scriptstyle \pm 0.3}$\\
        Latent PRM & $\mathbf{36.25}{\scriptstyle \pm 0.41}$ & $\mathbf{45.0}{\scriptstyle \pm 0.2}$ & $\mathbf{79.2}{\scriptstyle \pm 0.3}$\\
        \bottomrule
    \end{tabular*}
\end{table}

Table~\ref{tab:branch-selection} provides bounded support for the mechanism behind latent guidance.
Relative to random selection, the PRM more often selects a branch that improves over the original continuation or at least preserves it. While only modestly improving strict best-branch accuracy,
this local advantage is applied repeatedly across consecutive latent steps and multiple candidate branches, where choices can compound before decoding begins.
Together with the poor random-branch result in Table~\ref{tab:main-results}, this suggests that the PRM mainly helps by filtering harmful latent perturbations rather than by reliably finding the single best branch.

\FloatBarrier
\subsection{Ablations}
\label{sec:abilation}
\paragraph{Reward Construction Ablation.}
Table~\ref{tab:appendix-reward-ablation} compares the full mixed reward with grouped executable-only and auxiliary-only PRM supervision. All variants use the same latent-guidance inference configuration, $B_{\mathrm{test}}=8$ and $K_{\mathrm{test}}=12$, and each entry is the mean executable validation rate over three runs.

\begin{table}[t]
    \centering
    \small
    \begin{tabular*}{\columnwidth}{@{\extracolsep{\fill}}lr}
        \toprule
        PRM supervision  & Validation (\%) \\
        \midrule
        Full mixed reward & $42.10$ \\
        Executable signals only & $26.75$ \\
        Auxiliary semantic signals only & $29.38$ \\
        \bottomrule
    \end{tabular*}
    \caption{Three-run mean executable validation rate for the grouped PRM-supervision ablation on the ParaTrans test set.}
    \label{tab:appendix-reward-ablation}
\end{table}

 The executable-only target retains compilation, execution, and integrated pass/fail or output-based behavioral checks. The auxiliary-only target retains code-embedding similarity, the LLM judge, and the variable-comparison signal. For each restricted variant, the active weights are renormalized over the retained signals. The full mixed target performs best, reaching 42.10\% validation compared with 26.75\% for executable-only supervision and 29.38\% for auxiliary-only supervision.

This grouped ablation does not isolate every validator and therefore does not establish that each component is necessary. It directly supports only the broader choice of combining executable and auxiliary supervision. The result is consistent with binary executable feedback being too sparse by itself and semantic proxies being insufficient on their own for executable correctness. The reported task metric remains compilation, execution, and integrated validation against the reference behavior.

\paragraph{Latent Search Budget Ablation.}
Table~\ref{tab:appendix-bk-ablation} ablates the test-time latent search budget by varying the number of candidate branches per step, $B_{\mathrm{test}}$, and the number of latent reasoning steps, $K$. The $B_{\mathrm{test}}=1$ rows are unguided latent-reasoning controls; rows with $B_{\mathrm{test}}>1$ use learned PRM selection. Each entry is the mean executable validation rate over three runs.

Due to time and compute constraints, we construct PRM training trajectories under the smaller latent-step budget, but the preliminary pattern suggests that the PRM benefits from a larger decision surface at inference. Although task-specific PRM trajectories end at $K_{\mathrm{train}}=6$, the adapter, Qwen backbone, and value head share parameters across positions, and the backbone was pretrained on substantially longer sequences. 
More reasoning steps and more candidate options give the scorer additional opportunities to avoid weak continuations and occasionally choose stronger latent branches before decoding.
We treat $K_{\mathrm{test}}=12$ as an empirical inference-time budget extension rather than guaranteed length generalization.

\begin{table}[t]
    \centering
    \small
    \begin{tabular*}{\columnwidth}{@{\extracolsep{\fill}}lcc}
        \toprule
        $B_{\mathrm{test}}$ & $K$ & Validation (\%) \\
        \midrule
        1 & 6 & $26.32$ \\
        3 & 6 & $28.94$ \\
        1 & 12 & $32.89$ \\
        3 & 12 & $35.52$ \\
        8 & 12 & $\mathbf{42.10}$ \\
        \bottomrule
    \end{tabular*}
    \caption{Three-run mean validation rate (\%) for the test-time latent-search-budget ablation on the ParaTrans test set.}
    \label{tab:appendix-bk-ablation}
\end{table}

\clearpage
\section*{Limitations}

This study is limited to one base-model family, one PRM backbone, and one executable parallel-code translation benchmark. The ParaTrans test set contains 76 held-out tasks, so the results should be read as evidence for this setting rather than as a general claim about latent guidance across models or domains. 
We also do not evaluate transfer of the learned value model across base models. 
The adapter architecture and perturbation scale remain fixed, while the three-attempt repair budget is controlled but not ablated. We also do not compare greedy selection with beam-style latent search, which would alter the test-time compute budget and introduce additional search hyperparameters.
We leave these broader evaluations to future work.

Latent PRM guidance assumes access to intermediate hidden states, which limits its application to self-hosted models or otherwise white-box models unless black-box APIs expose these states.

As with other domain-specific adaptation methods, the approach also requires task-specific supervision to train the guidance model.
In our setting, this supervision is costly because parallel-code translations are long and each labeled example requires multiple rollouts followed by executable and semantic validation. This limits the scale of the present study and makes broader robustness and transfer evaluations more difficult.

Finally, although the branch-selection analysis is consistent with the filtering interpretation, it does not by itself explain all success and failure cases. Finer-grained analysis of these cases could inform more reliable latent intervention strategies beyond local branch filtering.

\section*{Risk Statement}

We do not identify material new risks from this work. The study is limited to offline benchmark evaluation of parallel-code translation.

\section*{Acknowledgments}

Computational support was provided by Code Metal.

\bibliography{bibliography}

\appendix

\section{Expanded Related Work and Positioning}
\label{sec:appendix-related}

\begin{table*}[t]
    \centering
    \footnotesize
    \begin{tabular*}{\textwidth}{@{\extracolsep{\fill}}p{1.4cm}p{2.8cm}p{1.8cm}p{1.6cm}p{1.3cm}p{3.9cm}}
        \toprule
        Area & Representative work & Intervention point & Representation & Exec.? & Difference \\
        \midrule
        Code generation and translation & \citet{chen2021evaluating}, \citet{chen2022codet}, \citet{ahmad2023summarizegenerate}, \citet{huang2023programtranslation}, \citet{bitan2025unipar} & Training data, decoded programs, or repair loops & Tokens or program text & Often & We intervene before final decoding by scoring latent prefixes while keeping the generator frozen. \\
        Parallel and HPC translation & \citet{nichols2024hpc}, \citet{chaturvedi2025hpc}, \citet{dearing2024lassi}, \citet{kaplan2026paracodexprofilingguidedautonomouscoding}, \citet{mueller2024syclomatic}, \citet{rocm2023hipify}, \citet{Martinez2011CU2CLAC}, \citet{shi2025transcl} & Fine-tuning, source-to-source translation, or post hoc repair & Program text or symbolic code & Often & These methods operate on decoded code or deterministic migration rules rather than on continuous latent prefixes. \\
        PRMs and verifiers & \citet{cobbe2021trainingverifierssolvemath}, \citet{lightman2023let}, \citet{wang2024math}, \citet{li2025codeprm}, \citet{dai2024process} & Token steps or completed programs & Text & Often & We supervise hidden-state prefixes before code is emitted. \\
        Latent reasoning and latent guidance  & \citet{hao2025traininglargelanguagemodels}, \citet{xu2025softcot}, \citet{you2026parallel}, \citet{yue2025hybrid}, \citet{zhang2025soft}, \citet{deng2025latent}, \citet{macfarlane2025searchinglatentprogramspaces}, \citet{zou2025latentcollaborationmultiagentsystems} & Internal reasoning phase or latent search & Hidden states or latent programs & Usually no & We attach a smaller PRM to score candidate latent branches for executable parallel-code translation. 
        We study execution-supervised prefixes for non-unique code translation, selecting discrete branches. \\
        \bottomrule
    \end{tabular*}
    \caption{Taxonomy of the closest neighboring literatures by intervention point.}
    \label{tab:appendix-taxonomy}
\end{table*}

Execution-based and compiler-guided code-generation work has established that behavioral tests are a stronger signal for program quality than string overlap~\citep{chen2021evaluating, chen2022codet}. Program-translation work then extended that view to translation settings through supervised adaptation, back-translation, distillation, and dialogue-based data generation~\citep{ahmad2023summarizegenerate, huang2023programtranslation, chen2025beyond}. These papers are relevant because they define the broader code-generation landscape in which our task sits. 
% but their intervention points remain at the level of training data or decoded text rather than pre-decoding latent guidance. 
These compiler-feedback and executable-repair methods compile, execute, rank, or revise decoded programs. Our method instead uses those delayed outcomes to train a scorer that acts before final code decoding; the shared repair-loop comparison tests whether the pre- and post-decoding intervention points are complementary.

HPC and parallel language translation-specific work is closer to our evaluation domain. UniPar, LASSI, ParaCodex, HPC-Coder, and HPC-Coder-v2 study executable parallel-code generation, while adjacent work studies parallelization assistance, fine-tuning, repair, or agentic adaptation with LLMs~\citep{nichols2024hpc, bitan2025unipar, dearing2024lassi, kaplan2026paracodexprofilingguidedautonomouscoding, chaturvedi2025hpc, harel2023learning, schneider2023mpi, kadosh2023advising, harel2025pragformer, bolet2025can, kadosh2024monocoderdomainspecificcodelanguage, schneider2024mpirigenmpicodegeneration, chen2024ompgpt, bolet2025counting}. Source-to-source translation systems such as SYCLomatic, HIPIFY, cu2cl, and TransCL intervene through deterministic or compiler-style program transformation rather than latent guidance~\citep{mueller2024syclomatic, rocm2023hipify, Martinez2011CU2CLAC, shi2025transcl, kadosh2024ompar, harel2020source}. Our setting is complementary: we ask whether a frozen generator can be steered before code is decoded. Thus, it is independent of downstream repair and transformation methods and
can be combined with downstream repair or optimization methods.

Most existing guidance methods intervene after the program text is produced. Verifier-guided reasoning, chain-of-thought prompting, search, and PRMs provide the conceptual template for allocating additional test-time computation with auxiliary signals~\citep{wei2022chain, nye2021show, cobbe2021trainingverifierssolvemath, lightman2023let, wang2024math, yao2023tree, zhang2025lessons}. For code, CodePRM and process-supervision-guided policy optimization show that execution-aware supervision can improve program generation~\citep{li2025codeprm, dai2024process}.
The key difference is that these methods score textual reasoning steps, decoded programs, or execution feedback after the text has been produced, whereas our method intervenes earlier by scoring continuous latent prefixes before code is emitted.

% Latent reasoning offers an opportunity for this different intervention point. These papers motivate continuous hidden states as a useful reasoning substrate and shift part of test-time computation from explicit token sequences into continuous hidden-state trajectories before returning to standard decoding~\citep{hao2025traininglargelanguagemodels, xu2025softcot, yue2025hybrid, zhang2025soft, deng2025latent, zou2025latentcollaborationmultiagentsystems}. ~\citet{macfarlane2025searchinglatentprogramspaces} is especially relevant because it treats latent computation as a search problem.
Latent reasoning offers an opportunity for this different intervention point. Prior work motivates continuous hidden states as a useful reasoning substrate and shifts part of test-time computation from explicit token sequences into continuous hidden-state trajectories before returning to standard decoding~\citep{hao2025traininglargelanguagemodels, xu2025softcot, yue2025hybrid, zhang2025soft, deng2025latent, zou2025latentcollaborationmultiagentsystems}. \citet{macfarlane2025searchinglatentprogramspaces} is especially relevant because it treats latent computation as a search problem.
\citet{you2026parallel} is closest methodologically, combining stochastic latent sampling with LatentRM-guided best-of-$N$ and beam search. It learns from rollout correctness on mostly arithmetic tasks. Instead we train a separate PRM from executable and semantic feedback for non-unique code translation and guide a frozen generator before decoding in a domain that requires task-specific expertise.
Concurrent work by \citet{li2026latentrewardsteering} uses reward gradients to optimize fragile sparse-autoencoder states during token-level reasoning, while we instead select discrete branches in a dedicated latent phase before decoding.
A more distant but relevant line of work is activation steering.
These methods modify internal activations at inference time, often by adding a contrastively derived direction to control a specific behavior \citep{turner2023steering, rimsky2024steering}.
Code-focused variants use similar interventions for type prediction or language and library control \citep{lucchetti2025understanding, rahman2026steering} but not in parallel code.
We draw only on the broader idea that hidden vectors can serve as intervention points. Unlike activation steering, however, we move the intervention from decoding-time control to the latent-reasoning phase, and we use a learned value model to compare task-conditioned continuations rather than applying a fixed steering direction.

\section{Supporting Results}
\label{sec:appendix-supporting-results}
\paragraph{Paired task win/loss/tie analysis.}
\label{sec:appendix-win-loss-tie}
We report task-level paired win/loss/tie counts using ParaTrans tasks averaged over three runs, matching the unit used for the paired bootstrap analysis. For each task, we compute the guided-minus-unguided validation-rate difference. A task is a win if the difference is positive, a loss if it is negative, and a tie otherwise.
Confidence intervals and p-values follow the task-level paired bootstrap and two-sided sign-flip tests described in Appendix~C, using 100{,}000 resamples or Monte Carlo samples.
Table~\ref{tab:paired-wins-app} shows that most tasks are unchanged, but among tasks whose outcome changes, PRM guidance more often improves than degrades validation in both the no-repair and repair settings.

\begin{table}[t]
\centering
\footnotesize
\begin{tabular*}{\columnwidth}{@{}l@{\extracolsep{\fill}}ccc@{}}
\toprule
Setting & Wins & Ties & Losses \\
\midrule
No repair & 20 & 50 & 6 \\
+3-attempt repair & 18 & 53 & 5 \\
\bottomrule
\end{tabular*}
\caption{Task-level paired win/loss/tie counts for PRM-guided latent reasoning against unguided latent reasoning on the 76-task ParaTrans test set.}
\label{tab:paired-wins-app}
\end{table}

\paragraph{Direction Split.}
Table~\ref{tab:appendix-direction} breaks down compilation and validation rates by translation direction. PRM guidance improves over unguided latent reasoning in three of the four directions, with the largest gain on Serial$\rightarrow$OpenMP, smaller gains on the two CUDA/OpenMP directions, and only compilation gain on Serial$\rightarrow$CUDA. We view this split as additional descriptive evidence that the aggregate result is not driven by only one direction, rather than as proof of direction robustness.
This pattern suggests that the aggregate improvement reflects both syntactic and semantic effects, since some directions improve mainly in compilation, some in validation, and some in both.

\label{sec:appendix-direction}

\begin{table}[t]
\centering
\footnotesize
\begin{tabular*}{\columnwidth}{@{}l@{\extracolsep{\fill}}cccc@{}}
\toprule
Method & C$\rightarrow$O & O$\rightarrow$C & S$\rightarrow$O & S$\rightarrow$C \\
\midrule
LR & 55.6/40.7 & 40.4/31.6 & 63.3/35.0 & 40.4/24.6 \\
LR-PRM & 53.7/46.3 & 47.4/36.8 & 65.0/60.0 & 42.1/24.6 \\
\bottomrule
\end{tabular*}
\caption{Direction-wise compilation/validation rates on ParaTrans. Abbreviations: LR = latent reasoning, C = CUDA, O = OpenMP, S = Serial.}
\label{tab:appendix-direction}
\end{table}

\paragraph{Test-time compute cost.}
\label{sec:appendix-test_time_compute}
We estimate the incremental cost of PRM guidance relative to unguided latent reasoning under the same $K=12$ latent-step setting with the 70B generator and one final decoding pass. PRM guidance adds $B_{\mathrm{test}}=8$ branch scores per step, including the unperturbed branch, or $8 \cdot 12 = 96$ Qwen-Coder-7B prefix-scoring calls per sample. By parameter count, this is roughly $9.6$ LLaMA-3.3-70B-sized forward-step calls, while avoiding additional 70B rollouts or extra decoded programs. This is only an approximate call-level normalization and does not account for implementation details. The decoded non-code portion is unchanged in our logs, averaging approximately 160 tokens in both settings, with high variance, and the final code is decoded once in each condition.

This overhead is small compared with post-decoding repair. A repair attempt requires another full 70B decoding pass over the program, which in our logs generates approximately 1,853 tokens per attempt on average. Thus, the three-attempt repair setting adds approximately $3 \cdot 1{,}853$ additional 70B decoding steps, compared with a much smaller auxiliary PRM-scoring budget. Despite this much smaller extra budget, latent PRM guidance without repair reaches 42.1\% validation, exceeding unguided latent reasoning with three repair attempts at 36.4\%. This comparison suggests that the gains are not explained simply by adding test-time computation.

\begin{table}[t]
    \centering
    \small
    \begin{tabular*}{\columnwidth}{@{\extracolsep{\fill}}lcc}
        \toprule
        Method & No repair & +3 repair \\
        \midrule
        Vanilla & $144.86$ & $399.89$ \\
        Fine-tuned & $144.52$ & $328.45$ \\
        Latent reasoning & $145.67$ & $315.93$ \\
        Random branch selection & $145.81$ & -- \\
        Latent PRM guidance & $147.33$ & $327.40$ \\
        \bottomrule
    \end{tabular*}
    \caption{Mean model-inference time per sample (seconds) for the Table~\ref{tab:main-results} configurations. Measurements use batch size 1 on two H200 GPUs. Compilation and validation, run separately on A40 GPUs, are excluded.}
    \label{tab:appendix-wall-time}
\end{table}

Table~\ref{tab:appendix-wall-time} reports mean model-inference time for every Table~\ref{tab:main-results} configuration.
We treat the small differences in inference time as run-to-run variation rather than evidence that any method is faster.
Times with repair additionally reflect the number and length of repair attempts triggered by failed translations and do not include the compilation and execution time for each kernel.

These measurements should not be read as a full efficiency claim. They show only that PRM scoring adds modest overhead in our implementation and remains compatible with token-level repair.

\paragraph{Post-decoding text-PRM reranking.}
We also evaluate a post-decoding best-of-8 control that applies the same reward supervision after complete programs have been generated. 
We sample eight complete outputs from unguided latent reasoning without latent perturbations, using temperature 0.8 during decoding, and train a text PRM over completed answers with the same reward function used for latent PRM training.
Candidates selected by the text PRM achieve 25.00\% validation, while  an oracle@8 over the same candidate sets reaches 36.84\%.
The oracle result indicates that the sampled sets sometimes contain validating programs, while the gap between oracle@8 and the text PRM shows that this control does not reliably identify them. Because the latent and post-decoding candidate sets are generated differently, this is not a matched compute or candidate-set comparison.
However, even this oracle remains below latent PRM guidance at 42.10\%. This suggests that the main gain is not explained by PRM supervision alone. Taken together, these results support the value of applying the reward model to latent prefixes before decoding in this setting.

\section{Hyperparameters and Reproducibility}
\label{sec:appendix-hparams}

\paragraph{Evaluation protocol.}
Main evaluation uses the 76-task ParaTrans test set, split across CUDA$\rightarrow$OpenMP (18), OpenMP$\rightarrow$CUDA (19), Serial$\rightarrow$OpenMP (20), and Serial$\rightarrow$CUDA (19). We report three-run mean executable validation rates with and without a three-attempt self-repair loop. In the repair setting, failed translations are regenerated from the previous code and the execution-environment feedback. We also report a separate 500-instance held-out branch-selection benchmark, where each saved latent prefix is paired with the original branch and two perturbed alternatives, and instances without a unique best branch are excluded.

For pairwise comparisons, we use task-level paired analysis. For each task, we average the binary validation outcomes over the three runs for each method, then compute the guided-minus-unguided task difference. We report the mean of these task-level differences. Confidence intervals are computed using a paired nonparametric bootstrap over tasks with $B=100{,}000$ resamples. P-values are computed using a two-sided paired sign-flip randomization test over the same task-level differences with $100{,}000$ Monte Carlo samples.

We used two NVIDIA H200 GPUs for primary-model inference when creating the dataset and running experiments. For PRM training and code evaluations, we used two NVIDIA A40 GPUs.
For the three-attempt repair setting, we feed the execution-environment error and the previously generated code back to the model, then ask it to regenerate the full code. The exact prompt templates are shown below.

\paragraph{PRM optimization.}
We train the PRM with the AdamW optimizer \citep{loshchilov2019decoupled} in two stages. In the first stage, we train only the adaptation layer, which maps the primary model hidden dimension of 8192 to the PRM backbone hidden dimension of 3584, together with the value head, for half an epoch. In the second stage, we unfreeze the remaining layers. We train with an effective batch size of 6 and a learning rate of $1\cdot 10^{-6}$.

\paragraph{Model selection.}
We selected LLaMA-3.3-70B for comparability with prior ParaTrans and UniPar-style evaluations and for compatibility with our COCONUT-style latent-reasoning implementation. We selected Qwen-Coder-7B as a smaller code-specialized auxiliary scorer, preserving a clear separation between the frozen generator and the learned PRM. We did not ablate generator or PRM size, so these choices should not be interpreted as necessary, sufficient, or optimal.

\paragraph{Noise-scale calibration.}
We set the perturbation standard deviation to $\sigma=1.4$ based on a preliminary two-point calibration on the 60-sample development split. With $B_{\mathrm{train}}=3$ candidates at each of $K_{\mathrm{train}}=6$ latent steps, the mean within-candidate reward range, $\max_b S(y_t^{(b)})-\min_b S(y_t^{(b)})$, averaged over samples and latent steps, was 0.060 for $\sigma=0.7$ and 0.086 for $\sigma=1.4$. The modest increase suggested greater separation among candidate outcomes, so we fixed $\sigma=1.4$ for subsequent experiments. This comparison is not a systematic sweep and does not establish that $\sigma=1.4$ is optimal.

\paragraph{Split and leakage audit.}
We define each translation task by the fingerprint
$(\texttt{kernel\_name}, \texttt{from\_api}, \texttt{to\_api})$.
The PRM development split contains the first 15 kernels in each
translation direction, for 60 examples total. It is used only for
hyperparameter selection. After hyperparameters are fixed, the final PRM
is trained on the union of the PRM training and development pools.
For the branch-selection experiment, we train a separate PRM after removing the 120 source trajectories used to construct the branch-choice instances.
The 76-task ParaTrans test set is held out from all PRM training,
development, trajectory collection, reward construction, and
hyperparameter selection. As a stricter check, we also compare kernel
names while ignoring translation direction; this overlap is also zero.

\begin{table}[t]
\centering
\small
\begin{tabular*}{\columnwidth}{@{\extracolsep{\fill}}lcc}
\toprule
Split / artifact & Count & Test-kernel overlap \\
\midrule
PRM dev & 60 & 0 \\
Branch-selection PRM train & 832 & 0 \\
Final PRM train & 952 & 0 \\
Branch-selection source set & 120 & 0 \\
ParaTrans test & 76 & -- \\
\bottomrule
\end{tabular*}
\caption{Leakage audit for PRM development/training artifacts and the held-out source trajectories used to construct the branch-selection benchmark. The branch-selection source set row counts trajectories, not the 500 derived branch-choice instances; overlap with the ParaTrans test set is computed using \texttt{(kernel\_name, from\_api, to\_api)} fingerprints.}
\label{tab:leakage-audit}
\end{table}

\paragraph{Reward construction.}
The reward target mixes executable and proxy components. We keep that distinction explicit because the benchmark metric in the main paper is executable validation only. For a terminal program $y$, let $C(y)\in\{0,1\}$ indicate successful compilation, $E(y)\in\{0,1\}$ indicate successful execution without runtime failure, and $A_j(y)\in[0,1]$ denote optional semantic or proxy validators available for that sample. We compute
\[
\resizebox{0.96\columnwidth}{!}{$
S(y)=0.30 C(y)+0.25 E(y)+0.45
\frac{\sum_{j\in\mathcal{A}(y)} w_j A_j(y)}{\sum_{j\in\mathcal{A}(y)} w_j}
$}
\]
where unavailable validators are omitted and the remaining validator weights are renormalized. The optional validators include pass/fail integrated validation, checksum or output matching when available, benchmark-specific metrics, Jina code-embedding cosine similarity to the reference target code \citep{günther2024jinaembeddings28192token}, a GPT-5-mini judge score, and the variable-comparison heuristic described below. In the common case where pass/fail validation, Jina similarity, and the GPT-5-mini judge are available, their effective contributions to $S(y)$ are $0.2423$, $0.1038$, and $0.1038$, respectively, in addition to the fixed $0.30$ compilation and $0.25$ execution weights.

\paragraph{GPT-5-mini judge.}
The judge is used only for PRM target construction and never for final benchmark scoring. For each candidate rollout, GPT-5-mini receives the source program, and the generated target-language program. It is asked to assess semantic equivalence, target-API correctness, and likely behavioral preservation, then return a bounded scalar score in $[0,1]$ with a short justification. We use the scalar score as one optional validator in the renormalized reward above.

\begin{promptbox}{Judge Prompt Template}
\small
\textbf{Role.} You are an expert in HPC code translation evaluation. Compare the original and translated code section below. Assign a score between 0.0 and 1.0 based only on syntax correctness, faithfulness, and overall quality.

\textbf{Scoring rubric.}
\begin{itemize}[leftmargin=1.35em,topsep=0.25em,itemsep=0.15em,parsep=0pt,partopsep=0pt]
    \item 0.00--0.20: incorrect or not on the right track
    \item 0.21--0.50: major issues
    \item 0.51--0.80: mostly correct with minor issues
    \item 0.81--1.00: strong and faithful
\end{itemize}

\textbf{Rules.}
\begin{itemize}[leftmargin=1.35em,topsep=0.25em,itemsep=0.15em,parsep=0pt,partopsep=0pt]
    \item Focus on logic and correctness.
    \item Ignore style, but pay attention to pragma placement in OpenMP directives.
    \item If the code is incomplete because of cut off text or missing includes, assume the most reasonable completion for scoring.
\end{itemize}

\textbf{Inputs.}
\begin{itemize}[leftmargin=1.35em,topsep=0.25em,itemsep=0.15em,parsep=0pt,partopsep=0pt]
    \item Kernel name: \promptplaceholder{\{kernel\_name or 'UNKNOWN'\}}
    \item Original code: \promptplaceholder{\{original\_code[:8000]\}}
    \item Translated code: \promptplaceholder{\{translated\_code[:8000]\}}
\end{itemize}

\textbf{Required output.} Return strict JSON only:
\begin{quote}
\small\ttfamily
\{"score": <float 0.0-1.0>, "reason": "<short reason>"\}
\end{quote}
\end{promptbox}

\paragraph{Variable-comparison heuristic.}
The variable-comparison heuristic is intended to catch semantic shortcuts not exposed by unit tests alone. GPT-5-mini first identifies variables in the source and generated programs that represent the core computed quantity for the built-in test. We then instrument the reference and generated programs by adding print statements for the selected variables, execute both programs on the same test input, and compare the observed final values.
\begin{promptbox}{Variable Identification Prompt Template}
\small
\textbf{Role.} You are a precise static code analysis agent.

\textbf{Hard rules.}
\begin{itemize}[leftmargin=1.35em,topsep=0.25em,itemsep=0.15em,parsep=0pt,partopsep=0pt]
    \item Static analysis only: do not suggest edits, do not modify code, and do not add instrumentation.
    \item Use only the code text provided in the request.
    \item Output must be valid JSON only, with no markdown or extra text.
\end{itemize}

\textbf{Input fields.}
\begin{itemize}[leftmargin=1.35em,topsep=0.25em,itemsep=0.15em,parsep=0pt,partopsep=0pt]
    \item \promptplaceholder{kernel\_name}: string
    \item \promptplaceholder{original\_code}: string, possibly empty, containing concatenated files prefixed by \promptplaceholder{// File: <filename>}
    \item \promptplaceholder{translated\_code}: string, possibly empty, using the same format
\end{itemize}

\textbf{Goal.} For each codebase, identify 1 to 3 variables that represent the most semantically central computed results of the program or kernel.
\begin{itemize}[leftmargin=1.35em,topsep=0.25em,itemsep=0.15em,parsep=0pt,partopsep=0pt]
    \item Return multiple variables only when there are genuinely multiple important result variables.
    \item If there is one clear result variable, return only that one.
    \item Order variables by importance, with the most central first.
\end{itemize}

\textbf{Critical distinction.}
\begin{itemize}[leftmargin=1.35em,topsep=0.25em,itemsep=0.15em,parsep=0pt,partopsep=0pt]
    \item Do not choose status flags, timing variables, loop counters, argument parsing variables, or configuration constants unless no computed result variable exists.
    \item Prefer numeric result variables and, failing that, the primary output buffer updated by the kernel or parallel region.
\end{itemize}

\textbf{Priorities.}
\begin{enumerate}[leftmargin=1.55em,topsep=0.25em,itemsep=0.2em,parsep=0pt,partopsep=0pt,label=\arabic*.]
    \item \textbf{Reported numeric result scalar.} Choose a non-boolean numeric scalar printed or logged as the main result, such as a checksum, error, norm, sum, energy, or output summary.
    \item \textbf{Derived numeric result variable just before output.} If the printed value is an expression, choose the stored numeric variable closest to the output site.
    \item \textbf{Primary output buffer, array, or pointer.} If there is no clear scalar result, choose the main computed output buffer.
    \item \textbf{Last resort.} If nothing else fits, choose the single variable most central to the computed output, avoiding booleans and timing variables when possible.
\end{enumerate}

\textbf{Type, missing, ambiguity, and not-found rules.}
\begin{itemize}[leftmargin=1.35em,topsep=0.25em,itemsep=0.15em,parsep=0pt,partopsep=0pt]
    \item Return the declared type exactly as written in code; use \promptplaceholder{UNKNOWN} only when the type cannot be determined.
    \item If \promptplaceholder{original\_code} is missing, set source status to \promptplaceholder{MISSING} with null fields.
    \item If \promptplaceholder{translated\_code} is missing, set target status to \promptplaceholder{MISSING} with null fields.
    \item If multiple candidates satisfy the same highest applicable priority, still choose one top candidate, mark the status as \promptplaceholder{AMBIGUOUS}, and list alternatives in notes.
    \item Use \promptplaceholder{NOT\_FOUND} only when code exists but no result variable can be identified under priorities 1--4.
    \item The \promptplaceholder{variables} array must contain 1 to 3 objects, ordered by importance.
\end{itemize}

\end{promptbox}

\paragraph{Inference prompt.}
The following prompt is used for the initial (non-repair) translation pass and during dataset creation. The system message establishes the model's role as an HPC translation expert, and the user template supplies the source API, target API, expected output file count and names, and the source code.

\begin{promptbox}{Regular Inference Prompt Template}
\small
\textbf{System.} You are an HPC expert specializing in translating between parallel programming APIs.

\textbf{User template.}
For each kernel code provided, think about it step by step, and then translate it from \promptplaceholder{\{from\_api\}} to \promptplaceholder{\{to\_api\}}.
The target \promptplaceholder{\{to\_api\}} implementation is spread across \promptplaceholder{\{n\}} files: \promptplaceholder{\{names\_str\}}.
Provide the complete code in \promptplaceholder{\{to\_api\}}.
Do not truncate or use ellipses. Do not change the main function.
Ensure correctness. All function names must match.
The code to translate:
\begin{quote}
\small\ttfamily
\{source\_code\}
\end{quote}
\end{promptbox}

\paragraph{Three-Attempt repair prompt.}
The repair prompt is used in the three-attempt regeneration setting after a failed compilation or execution attempt. It provides the previous error trace together with the broken translation and asks the model to emit a complete corrected translation in a strictly parseable format.

\begin{promptbox}{Repair Prompt Template}
\small
\textbf{Context.} The previous translation attempt \promptplaceholder{(iteration \{iteration - 1\})} failed with the following error:
\begin{quote}
\small\ttfamily
--- BEGIN ERROR ---\\
\{error\_text\}\\
--- END ERROR ---
\end{quote}

\textbf{Previous broken translation.}
\begin{quote}
\small\ttfamily
--- BEGIN PREVIOUS ATTEMPT ---\\
\{prev\_code\}\\
--- END PREVIOUS ATTEMPT ---
\end{quote}

\textbf{Instruction.} Please fix the error and produce a corrected translation.

\textbf{Output format requirements.}
\begin{itemize}[leftmargin=1.35em,topsep=0.25em,itemsep=0.15em,parsep=0pt,partopsep=0pt]
    \item Provide the complete code in \promptplaceholder{\{to\_api\}}.
    \item Do not truncate or use ellipses. Do not change the main function.
    \item Ensure correctness. All function names must match.
    \item Wrap the code in a single fenced block beginning with \verb|```cpp| and ending with \verb|```|.
    \item You may instead begin the block with \verb|```c|, \verb|```cuda|, \verb|```omp|, or \verb|```cu| if that better matches \promptplaceholder{\{to\_api\}}.
    \item If the translation spans multiple files, place a \promptplaceholder{// File: <filename>} marker at the start of each file's content inside the same fenced block, using the exact target filenames.
    \item Do not include any prose after the closing fence.
\end{itemize}
\end{promptbox}
\end{document}